\documentstyle[12pt]{article}
\begin{document}
\begin{center}
{\bf The Finite-Difference Analysis and Time Flow}\\[0.5cm]
Ashot Yu. Shahverdian\\[0.2cm]
\begin{small}
(Yerevan Physics Institute)\\[0.5cm]
\end{small}
\end{center}
{\bf 1.~~~Introduction} \\[0.2cm]
The present paper introduces a method of analysis of one
dimensional systems. Its application to study of Poincare
recurrence time flow is considered. The approach
suggested consists of reducing the research of a given system's
orbits ${\bar X}=(x_{i})_{i=1}^{\infty}$ to
analysis of alternations of the monotone increase and decrease
of higher order absolute finite differences, taken from
$\bar X$. Through some special representation of
finite orbits, we associate with ${\bar X}$ some
numerical sequence, is called the conjugate orbit.
We formulate two conditions, when the study of the
${\bar X}$ is reduced to analysis of the
special orbits' asymptotical intersections with
some base set from numerical interval $(0,1)$ of
zero Lebesgue measure. Hence, the approach
can be characterized as an "asymptotical" analogy
of Poincare's classical "section"method.

The method distinguishes two cases: the continuous,
when some numerical quantities $\rho$
can take arbitrary values from interval $(0,1)$, and
the discrete case, when all of them belong to a finite
set. The continuous was applied in Ref.~1 to study of neural
activity. In the present work, we are more
interested in the discrete case, which has an
additional specific. Thus, as soon as we have stated
the method's applicability, the mechanism generating time
series $\bar X$, is replaced by the return map $\cal R$,
generating the conjugate series. While the actual
exact description of this mechanism may remain unknown
(e.g., for earthquake time series or neuron spike trains),
the function $\cal R$ does not depend on it and has a simple
analytic shape.

This approach is applicable to the study of various time
series, arising in modern applied science (e.g., medicine,
astronomy). The innovation consists of a new way to
measure the fast oscillating time series -- this measure
is the Hausdorff dimension of some thin spaces
$\cal A$, to which the conjugate series are attracted.
As an example of such an application, the computational
analysis of the sequences of fractional parts
is presented.
We show that
the results obtained testify on the possibility of
Cantorian structure of our usual time flow. In this
context, we recall some well known (qualitative)
conclusions on irregular and intricate time flow,
as indicated in Bergson's philosophical treatments
\cite{hb} and in some works on quantum physics
\cite{ms,di}.\\[0.4cm]
{\bf 2.~~~Description of the Method} \\[0.2cm]
{\bf 2.1~~~Finite differences and conjugate orbits}\\[0.2cm]
We deal with one dimensional, deterministic or
stochastic systems, generating numerical sequences
${\bar X} = (x_{i})_{i=1}^{\infty}$ \
($0\leq x_{i}\leq 1$).
We impose no restrictions on the system,
and it may also possess different inner
states changing with time. Below, we set two conditions
on a given orbit $\bar X$,
under which it should be regarded as chaotic.
We introduce the notion of conjugate
(with $\bar X$) orbit $\bar \nu$, in terms of which
the irregular nature of $\bar X$ can be established
and studied.

Let us first give some special representation of
finite orbit ${\bar{X}}_{k}=(x_{i})_{i=1}^{k}$,
which reflects its higher order differential
structure. For the sequence ${\bar{X}}_{k}$
and number $1\leq s\leq k-1$ we let
  $$ \Delta_{i}^{(0)} = x_{i}, \qquad
  \Delta_{i}^{(s)} =
  |\Delta_{i+1}^{(s-1)} - \Delta_{i}^{(s-1)}|
  \quad (i=1, 2, \ldots, k-s). $$
It is not difficult to obtain
\begin{equation}
  \Delta_{i}^{(s-1)} = \mu_{k,s-1} +
  \sum_{p=1}^{i-1}(-1)^{\delta_{p}^{(s)}}
  \Delta_{p}^{(s)}  -
  \min_{0\leq i\leq k-s}(\sum_{p=1}^{i}
   (-1)^{\delta_{p}^{(s)}}\Delta_{p}^{(s)})
\end{equation}
where
$$
    \delta_{p}^{(s)} = \left \{ \begin{array}{ll}
    0 & \Delta_{p+1}^{(s)} \geq \Delta_{p}^{(s)} \\
    1 & \Delta_{p+1}^{(s)} <    \Delta_{p}^{(s)}
                                \end{array} \right.
    \quad \mbox{(it is supposed that} \quad
    \sum_{1}^{0} = 0) \ ,
$$
and
 $$ \mu_{k,s} =
 \min \{ \Delta_{i}^{(s)}: 1\leq i \leq k-s\} \ . $$
From Eq.~(1) it is easy to see, that if for a given $m$,
$0\leq m\leq k$ all of the finite binary sequences
$$
 \delta_{1}^{(s)}, \delta_{2}^{(s)}, \ldots,
 \delta_{k-s}^{(s)}
 \qquad (s = 0, 1, \ldots, m)
$$
as well as all of the numbers
$$
  \mu_{k,1}, \  \mu_{k,2}, \  \ldots , \mu_{k,m}
  \quad \mbox{and} \quad \Delta_{1}^{(m)},
  \Delta_{2}^{(m)}, \ldots, \Delta_{k-m}^{(m)}
$$
are known, one can completely restore the initial
finite orbit ${\bar X}_{k}=(x_{i})_{i=1}^{k}$. Hence,
it follows from (1), we can consider that finite
orbits with length $k$ are given namely in the next
special representation
\begin{equation}
{\bar \zeta}_{k} =
   (r_{1}^{(k)},r_{2}^{(k)},\ldots,r_{m}^{(k)};
   \mu_{k,1}, \  \mu_{k,2}, \  \ldots , \mu_{k,m};
   \rho_{1}, \  \rho_{2}, \  \ldots, \rho_{k-m})
\end{equation}
where
\begin{equation}
    r_{s}^{(k)} =
    0.\delta_{1}^{(s)}\delta_{2}^{(s)}\cdots
    \delta_{k-s}^{(s)} \qquad (1\leq s\leq m)
\end{equation}
are the rational numbers, and
$$
  \mu_{k,1}, \  \mu_{k,2}, \  \ldots ,
  \mu_{k,m}  \quad \mbox{and} \quad
  \rho_{1}, \  \rho_{2}, \  \ldots ,
  \rho_{k-m}  \quad
  (\mbox{where} \quad \rho_{i} = \Delta_{i}^{(m)})
$$
are some numbers from interval $[0,1]$. Here $m=m_{k}$
and $m_{k}$ tend to $\infty$ as $k\to \infty$.
It is easy to see, that after applying the recurrent
procedure (1) the sequence ${\bar{X}}_{k}$
is completely recovered by ${\bar{\zeta}}_{k}$.
Now, we can introduce the basic tool of the method -- the
notion of conjugate orbit: we say, that numerical sequence
${\bar \nu} = (\nu_{s})_{s=1}^{\infty}$ is the conjugate
orbit, associated with the given orbit
${\bar X}= (x_{i})_{i=1}^{\infty}$, if for each
$s\geq 1$ the terms  $\nu_{s}$ are defined as follows
$$
  \nu_{s} = \lim_{k\to \infty}r_{s}^{(k)} \qquad
  (= 0.\delta_{1}^{(s)}\delta_{2}^{(s)}
  \delta_{3}^{(s)}\ldots)
$$
were $r_{s}^{(k)}$ are the $r$-coordinates from (2).

Let us introduce a thin set $\cal B$ and its subsets
${\cal B}_{k}$, are the base sets, mentioned in Sec.~1.
We consider the numbers $0<x<1$ represented in
the form of binary expansion,
\begin{equation}
x=0.\delta_{1}\delta_{2}\delta_{3}\ldots  \qquad
  (=\sum_{k=1}^{\infty}2^{-k}\delta_{k}, \quad
  \delta_{k}=0, 1) \ ;
\end{equation}
further, we also operate with the corresponding
binary sequences $\bar x$,
 $${\bar x}=
 (\delta_{1}, \delta_{2}, \delta_{3}, \ldots,
 \delta_{n},\ldots) \ . $$
For a given natural $k\geq 2$ we define the subsets
${\cal B}_{k}$ of numerical interval $(0,1)$:
${\cal B}_{k}$ is the set all of those real numbers
$x\in (0,1)$ for each of which
$$
 n_{i+1}-n_{i} \leq k \qquad (i=1, 2, 3, \ldots)
$$
where $n_{i}$ denote all the consecuitive positions
where the changes of binary symbol from (4) occur,
$\delta_{n_{i+1}} = 1-\delta_{n_{i}}$. Let also
$\cal B$ be the union of all ${\cal B}_{k}$
($k\geq 2$). All of the sets ${\cal B}_{k}$, \
$ 2\leq k \leq \infty $, have zero Lebesgue measure
(see Ref.~1) (hereafter, in order to reduce some
formulations, we use also the notation
${\cal B}_{\infty} \  (\equiv {\cal B})$).

Let us now suppose, that the orbit $\bar X$ is such,
that each of the sequences
\begin{equation}
 {\bar X}^{s} = (\delta_{1}^{(s)}, \delta_{2}^{(s)},
 \ldots, \delta_{n}^{(s)}, \ldots)
 \qquad (s = 0, 1, \ldots)
\end{equation}
where $s$ is fixed, has bounded lengths of series
with the same binary symbol.
In other words, we assume that for increasing
sequence of indeces $n_{i}^{(s)}$,
designating all of those positions in natural series,
where the changes of binary symbol occur,
$\delta_{n_{i+1}^{(s)}}^{(s)} =
1-\delta_{n_{i}^{(s)}}^{(s)}$,
we have
\begin{equation}
  n_{i+1}^{(s)}-n_{i}^{(s)}\leq K_{s}
    < \infty \qquad (i = 1,2,\ldots) \ .
\end{equation}
Hence, taking both numbers $k$ and $k-m$
large enough (for determinity it can be
chosen $m$ is equal to entire part of
$k/2$, $m=[k/2]$), we can consider each of the
sequences (5) as the binary expansion of some number
$\nu_{s}\in {\cal B}_{K}$,
$$
  \nu_{s} = 0. \delta_{1}^{(s)} \delta_{2}^{(s)}
  \ldots \delta_{n}^{(s)}\ldots
$$
or, in other words, each of the sequences
$(r_{s}^{(k)})_{s=1}^{\infty}$
of the form (3) converges (as $k\to \infty$) to a number
$\nu_{s}$ from  ${\cal B}_{K}$. Here,
$2\leq K \leq \infty$, and according to (6) it should
be taken $K = \sup \ \{ K_{s}: s\geq 1 \}$. \\[0.15cm]
{\em Continuous Case}.~~Now, we asume also, that the quantities
$$
      \mu_{k} =  \sum_{i=1}^{m_{k}}\mu_{k,i}^{2} +
                 \sum_{i=1}^{k-m_{k}}\rho_{i}^{2}
$$
converge  to zero,
\begin{equation}
  \lim_{k\to \infty}\mu_{k} = 0  \ .
\end{equation}
In such a way, the limitations we impose on the time series $\bar
X$ are the following:
\begin{quote}
$(C_{1})$:~~~for every $k\geq 1$ the sequence
${\bar X}^{k} \in {\cal B}$
\end{quote}
\begin{quote}
$(C_{2})$:~~~the quantities $\rho_{i,k}$ and $\mu_{i,k}$ are
such, that: $\mu_{k}=o(1)$ as $k\to \infty$.
\end{quote}
Then, one can see \cite{av},
these rwo restrictions imply that the
transformed sequence ${\bar \zeta}_{k}$ from (2)
tends to space
${\cal B}^{\infty} \ (={\cal B}\times{\cal B}\times\ldots)$,
\begin{equation}
     ||{\bar{\zeta}}_{k} - {\cal B}^{\infty}|| = o\:(1)
  \qquad (k \to \infty)
\end{equation}
where $||.||$ is the usual metric in Euclidean spaces $R^{k}$, \
$||x|| = (\sum_{i=1}^{k}x_{i}^{2})^{1/2} $.
Particularly, (8) implies that for every
$s$-th ($1\leq s\leq m$) coordinate of ${\bar{\zeta}}_{k}$
we have
  $$\lim_{k\to\infty}{\bar{\zeta}}_{k}^{(s)} =
  \lim_{k\to\infty}r_{s}^{(k)} =
    \nu_{s} \in {\cal B} $$
-- namely this relation was meant when we mentioned the
"asymptotical" intersections (of transformed into a
special form orbits) with a thin space. \\[0.15cm]
{\em Discrete Case}.~~In this case we deal with one dimensional
time series with upper bounded terms and
with restricted measurement accuracy, known in advance.
If it is some number of the form $10^{-m}$, $m\geq 1$, then
multiplying all of the terms of our series by this quantity, we
will obtain some time series $\bar X$, consisting of natural numbers
bounded by a pregiven $N$,
\begin{equation}
{\bar X}= (n_{1}, n_{2}, n_{3}, \ldots, n_{k}, \ldots),
\qquad n_{k}\in [0, N]  \quad (k = 1, 2, 3, \ldots) \ .
\end{equation}
In other words, we deal with the random sequences \cite{me};
particularly, such kind of sequences arise as
a result of the realizations of hazard games.
However, we consider only a special class of
sequences (9) -- as above, we impose two
limitations (below $d\geq 1$ is some
natural number; we may assume $d=1$) on the differential
structure of the series $\bar X$:
\begin{quote}
$(D_{1})$:~~~for every $k\geq 0$ the sequence
${\bar X}^{k}\in {\cal B}_{K}$
\end{quote}
\begin{quote}
$(D_{2})$:~~~for all large enough $k\geq 1$ and all
$i\geq 1$ the quantities $\rho_{i}^{k} \in \{0, d\}$.
\end{quote}
It should be noticed, that condition $(D_{1})$ (and $(C_{1})$)
implies the dimensionality limitations (see next section), while
the $(D_{2})$, an analogy of $(C_{2})$, is a requirement on
certain regularity of the process, generating $\bar X$.

The restriction $(D_{2})$ means, that
we consider such processes (9), which are reduced to binary
processes $\bar X$ (with components $x_{i}=0, 1$).
Let us call the binary sequence $\bar X$, satisfying the
conditions $(D_{1})$ and $(D_{2})$ as $\beta_{K}$-sequence
and denote ${\hat{\cal B}}_{K}$ the collection all of the
numbers $x\in (0,1)$ for which its binary expansion $\bar x$
is ${\beta}_{K}$-sequence.
From the definition, we have the next characteristic
of these sets; below
$x_{n}=\Delta^{(n)}(x)$ and $\Delta$ is the shift
transformation (or the "tent map", see Ref.~6):
 $$ \Delta(x) = \left \{ \begin{array}{ll}
   2x   \ & 0<x<1/2 \\
   2x-1 \ & 1/2 \leq x <1 \ .
   \end{array} \right. $$
\newtheorem{guess1}{Proposition}
\begin{guess1}
The number $x\in {\hat {\cal B}}_{\infty}$ iff for some
numbers $0<K_{p}<\infty$ the relation
\begin{equation}
 |x_{n}-\Delta^{(p)}(x_{n})| \geq 2^{-K_{p}}
\end{equation}
holds for every $n\geq 0$ and $p\geq 1$. The number
$x\in {\hat {\cal B}}_{K}$, \ $2\leq K <\infty$, iff
the inequality (10) holds with $K_{p}\equiv K$ and
for every $n\geq 0$ and $p\geq 1$.
\end{guess1}
According to Poincare recurrence theorem (see, e.g., Ref.~6),
the relation
 $$ {\lim\inf}_{p\to \infty} |x-\Delta^{(p)}(x)| = 0  $$
holds for each $0<x<1$ except some set of zero Lebesgue
measure.
In the case $K_{p}\equiv K$ the condition (10) means recurrence
break (by analogy with "ergodicity break" from Ref.~7)
for all points $x_{n}$. Hence, all the sets
${\hat{\cal B}}_{k}$ are found within exceptional set,
mentioned in just quoted Poincare theorem.
Note also certain resemblance of the relation (10) with the
basic inequalities in Li - Yorke known theorems
\cite{ly}.\\[0.3cm]
{\bf 2.2~~Conjugate attractor and return map}\\[0.2cm]
Let us now denote ${\cal A}$ the closure of conjugate orbit
$(\nu_{s})_{s=1}^{\infty}$ , i.e. ${\cal A} \subset [0,1]$
is the union of the set $\{\nu_{s}: s\geq 1\}$ with the
collection of all its cluster points. Then this orbit
should be considered as a chaotic one, whenever
$\cal A$ fills in either an interval or a Cantor set.
We note, that for the most actual systems the
constants from Eq.~(6) for all $s\geq 1$ are upper bounded
by the same number $K\geq 2$ and, consequently,
\begin{equation}
  {\cal A}\subset {\cal B}_{K} \quad \mbox{and} \quad
  dim ({\cal A}) \leq dim ({\cal B}_{K}) \ .
\end{equation}
Hence, just mentioned criterion of chaosity simply means
that $\cal A$ is a Cantor set. If $\cal A$ appears to be
the same set for almost each orbit $\bar X$, then we can
conclude that $\cal A$ is the attractor of the system.
The numerical values of Hausdorff dimension of the sets
${\cal B}_{k}$ ($2\leq k\leq \infty $) can be obtained
from the formula \cite{av}
$$
z_{k}(2^{dim({\cal B}_{k})}) = 1
\quad \mbox{where} \quad
 z_{k}(s)=\sum_{n=1}^{k}s^{- n} \qquad (s>0)
$$
-- hence, (11) implies an upper estimate of Hausdorff
dimension of attractor $\cal A$.

If we consider the numbers $x$ from unit interval are
given in the form (4),
then it is not difficult to obtain the next statement
(below, $a\oplus b = |a-b|$ is the  logical
sum of the binary symbols $a, b \in \{0, 1\}$):
\begin{guess1}
If $\bar X$ is a binary sequence, then the return map
${\cal R}: \nu_{k}\to \nu_{k+1}$ of the conjugate
orbit $\bar\nu$ has the form
$${\cal R} (x) = \sum_{n=1}^{\infty}
2^{-n}(\epsilon_{n} \oplus  \epsilon_{n+1})$$
where $\epsilon_{n}$ are the coefficients of binary expansion
of number $x$.
\end{guess1}
One can find (in the operator form) the function $\cal R$
in Sinai's early constructions on weak isomorphism
of dynamical systems \cite{pb,yg}. For more details
on this function see Ref.~9, where a fractal dynamical
system is introduced.\\[0.4cm]
{\bf 3.~~~Sequences of Fractional Parts and Time Flow}\\[0.2cm]
Some works in quantum mechanics \cite{ms}
introduce the concept "Cantorian
Space-Time". The computational study of the flow of our
usual (but in some sense discretized) time, implemented in
this section, confirms again the possibility of such a concept.
We are based on the method introduced
-- an approach completely different from those considered
in other works on time flow \cite{rh,mf,gm}.

Let us start from the statement and
computational analysis of the
following number-theoretical problem. Let us have  two numbers
$0<\alpha, \lambda <1$, $\alpha $ is irrational and  let
\begin{equation}
 \{ \alpha \}, \  \{2\alpha\}, \  \{3\alpha\},
 \ldots, \{n\alpha\}, \ldots
\end{equation}
be the sequence of the fractional parts, generated by $\alpha$
(hereafter $\{x\}$ denotes the fractional part of number $x$).
We are interested in the behavior of the time series
  $${\bar X} = (n_{k})_{k=1}^{\infty} \quad \mbox{where} \quad
  n_{k}=m_{k+1}-m_{k} \qquad (k\geq 1)  \ ; $$
here $m_{k}$ are all of the natural numbers,
arranged in increasing order, $m_{k+1}>m_{k}$, such that
\begin{equation}
  \{ \alpha m_{k} \} \in (0,\lambda) \ .
\end{equation}
According to H.~Weyl theorem \cite{pb,hn} the sequence
(12) is uniformly distributed in the interval $(0,1)$, so that
we have infinite number of $m_{k}$, satisfying (13).
If we take into attention the ergodicity of the transformation
$\pi$,
\begin{equation}
  \pi(x)=\{ x + \alpha \} \ , \qquad \pi: (0,1) \to (0,1) \ ,
\end{equation}
this statement can also be deduced (for a.e.
irrational $\alpha$) from Poincare's recurrence theorem.
From numerous theoretical results on the sequences of
fractional parts (see details in Ref.~14), we note the following
Slater-Florec theorem: for given irrational
$\alpha$ there exist two natural numbers $p$ and
$q$, such that for all $k\geq 1$ we have
             $ n_{k}\in \{p, q, p+q\} $.
We may also note a formula, obtained in
Ref.~15 -- it expresses the exact value of the total number
of indeces $m_{k}$, satisfying (13) and not exceeding a given
$N$, through the coefficients of some expansions of the
numbers $N$ and $\lambda$.

Our basic conclusion in the problem stated, deduced from
computational experiments,  can be  formulated in the next form:
for each irrational $0< \alpha <1$ there exist the numbers
$0<\lambda <1$ such that the conjugate to $\bar X$ orbit $\bar\nu$ is
asymptotically close to a Cantor set; and conversely, for each
$0< \lambda <1$ there exist irrational numbers $0<\alpha <1$, such
that the conjugate to $\bar X$ orbit $\bar\nu$ is asymptotically
close to a Cantor set.

Some of these computational results, on
the Fig.~2 are presented. Here, the numerical value
$\lambda=0.13$ and several values of parameter $\alpha$ are
considered. In dependence of control parameter $\alpha$
the conjugate orbits $\bar\nu$ demonstrate all possible types of
motion: the periodic, when they are attracted to some finite set,
Cantorian, when the attractor $\cal A$ has Cantorian self-similar
structure, and the completely chaotic motion, when the orbits
fill in an interval from $(0,1)$. For instance, for
$\alpha=0.1250002$ the Cantorian attractor in Fig.~2.b is
shown. The graph Fig.~2.a has
been obtained for the value $\alpha = 0.124999$; in this case
the orbit $\bar\nu$ consists of 13 straight lines (some of them
are not shown).
In some respect, the behavior (in dependence of
control parameter
$\alpha$) of conjugate orbits in this problem reminds
Feigenbaum's transitions (see e.g., Ref.~6), and this needs
further detailed computational study.

To end, let us consider the following interpretation of
the above introduced number-theoretical problem; in particular,
this will explain the title of present section.
Let us have in a two dimensional plain a circle $E$ and
a point $S$, situated
outside of disc, bounded by $E$. Let $\omega$ be the angle,
under which the circle $E$ from the point $S$ is seen, and let
also $L$ be an arc of $E$ with the length
$|L|=\lambda <\omega$ with the center
at the point of intersection of circle $E$ with the rectilinear
segment, connecting its center with $S$.
Let us assume that the point $M\in E$ is rotated in discrete time
$T = \{ 1, 2, 3, \ldots. \}$ with the frequency $\alpha$. Then,
the consecuitive moments of falling the point $M$ in the arc $L$
constistute a sequence of natural numbers $m_{k} \in T$,
coinciding with the sequence $\bar X$ from Eq.~(13). Also,
it is not
difficult to see, that the sequence of fractional parts (12)
coincides with the sequence of iterates of the function
$\pi (x)$ from Eq.~(14), i.e.
 $$ \{\alpha n \} = \pi^{(n)}(\alpha)  \qquad (n\geq 1) $$
and, consequently, the numbers $m_{k}$ represent  the
Poincare recurrence (of point $M$ into the arc $L$) time.
Now, let us interpret $E$ as "Earth" and  $S$ as "Sun". Then,
replacing the continuous rotation of Earth by discrete motion,
we obtain that $m_{k}$ appears to be an approximate of the
flow of our usual time. In such a way, the above described
computations demonstrate, that in principle, it is possible
that our time
possesses the Cantorian structure. The numerical values of the
quantities $\alpha$, $\lambda$, and $\omega$ can be specified
for this "astronomical" case, but here we do not dwell on
this point. \\[0.4cm]
{\bf 4.~~~Some Remarks and Discussions}\\[0.2cm]
Let us discuss two problems, relating to the brain
activity. In the paper Ref.~1, through the method
above, applied to experimental data of actual neurons,
the Cantorian strucuture of brain activity has been
deduced. The method permits further development
\cite{ay} by means of introducing some dynamical
system ${\cal F} = ({\cal A}, T, \mu)$
on the Cantor spaces $\cal A$. The evolution
operator $T$ is easly expressed through the return map
$\cal R$, while the attractor $\cal A$ and invariant
(singular) measure $\mu$ remain the main quantities,
distinguishing different processes. In practical
applications, the measure $\mu$ can be
calculated through frequency analysis,
accepted in mathematical-linguistic works (see, e.g. Ref.~5).
Hence, we can operate with
the theoretical-information characteristics of system
$\cal F$. Whether there exists an analogy of known
Billingsley-Eggleston formula \cite{pb}
      $$Dimension = Entropy,$$
relating to Eggleston thin spaces, for the fractal
dynamical systems $\cal F$? Note also Mandelbrot's
work \cite{bm}, who considered the applications of
this formula to problems of turbulence.

We know \cite{ep}, that there exists some mechanism of time
perception in the brain. Its activity, it follows from our
previous results \cite{av}, must be found in some thin
(multidimensional) Cantor space.
How to relate the measurable spaces ${\cal F}_{t}$ and
${\cal F}_{p}$, which correspond to time flow and time
perception, with each other? Particularly, can we compare
the values of $dim ({\cal A}_{t})$ and $dim ({\cal A}_{p})$?

It seems, these problems can be resolved after detailed
computational frequency analysis,
in order to compute the invariant measures $\mu$ (and
hence another characteristics, say, entropy) of both time
flow and brain activity. \\[0.4cm]
{\bf Acknowledgements}\\[0.2cm]
The author is grateful to Dr. A. V. Apkarian from State
University of New York Health Science Center,
Syracuse, N-Y, USA, for support
and discussions of main topics of this work.

\newpage
\begin{center}
{\bf Figure Captions} \\[0.4cm]
\end{center}
Figure 1.~~The graph of inverse function
${\cal M}(x) = {\cal R}^{-1}(x)$. The graph is constructed
through 15,000 points, computed by means of random numbers
generator. From definition of the function
$\cal R$ it is easy to see that $\cal M$ is a
multivalent function, defined on the interval $(0,1)$.
The graph of the return map $\cal R$ can be obtained as a
result of symmetric reflection of this "fractal letter $M$"
in respect of diagonal $(0,0)$, $(1,1)$. \\[0.4cm]
Figure 2.~~Two different types of conjugate orbit's behavior
in dependence of control parameter $\alpha$. The
graph a) demonstrates the case when
the attractor $\cal A$ consists of the finite set, while in
the graph b) these orbits are attracted to some Cantor set.
The graph c) shows the self-similar structure of the Cantor
set $\cal A$ from the previous graph.
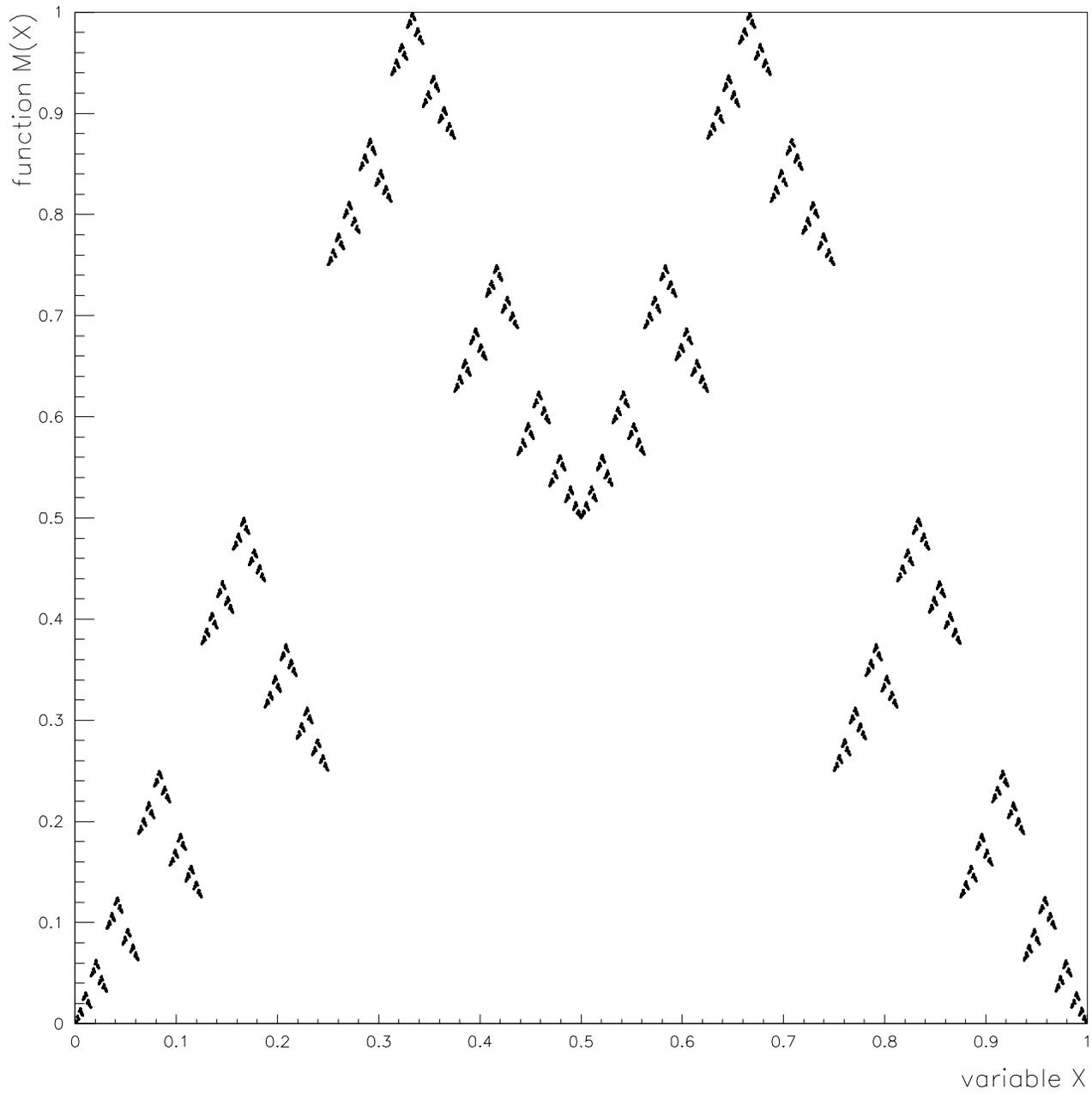
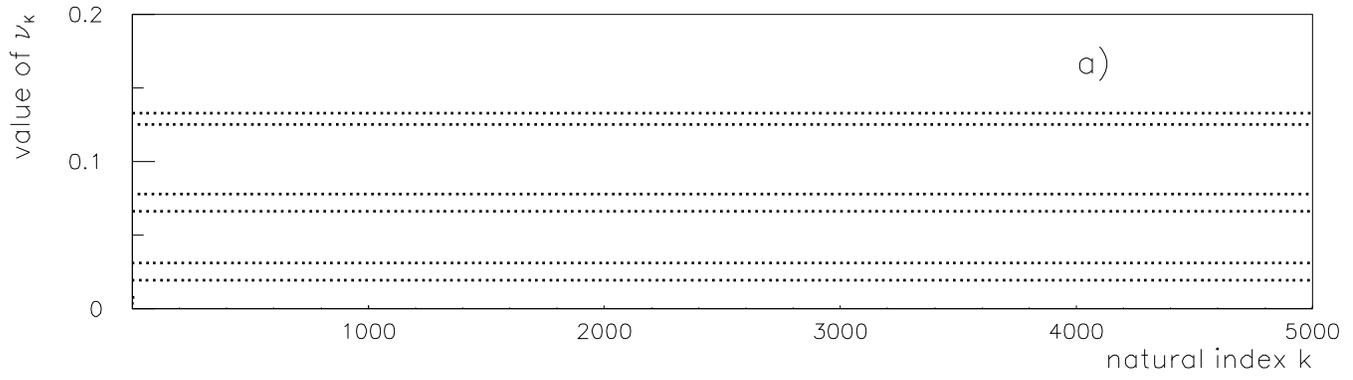
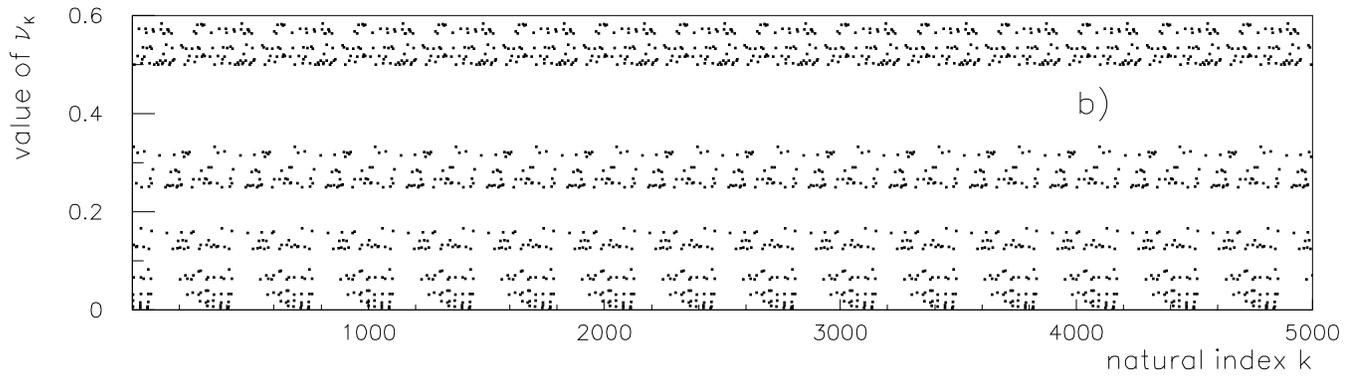
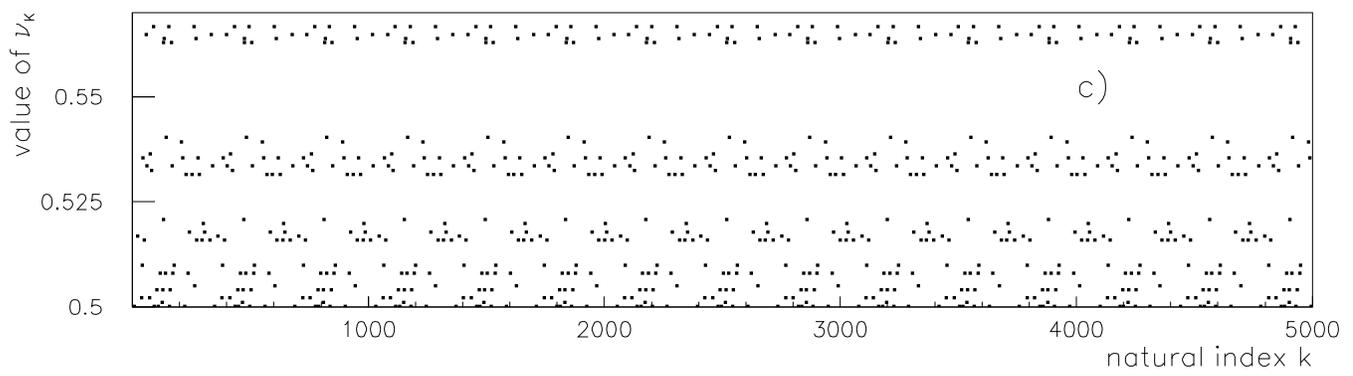
\end{document}